\documentclass[twocolumn,twoside,slac]{revtex4}
\usepackage{graphicx}
\usepackage{fancyhdr}
\begin{document}

\title{Reconstruction and Analysis on Demand: A Success Story}

%

\author{C. D. Jones}
\affiliation{Cornell University, Ithaca, NY 14853, USA}

\begin{abstract}
The traditional design of an HEP reconstruction system partitions the problem into a series of modules.  A reconstruction job is then just a sequence of modules run in a particular order with each module reading data from the event and placing new data into the event.  The problem with such a design is it is up to the user to place the modules in the correct order and CPU time is wasted calculating quantities that may not be used if the event is rejected based on some other criteria.

The CLEO III analysis/reconstruction system takes a different approach: on demand processing (otherwise known as lazy evaluation).  Jobs are still partitioned into smaller components which we call Producers.  However, Producers register what data they produce.  The first time a datum is requested for an event the Producer's algorithm is run. Sources work similarly, registering what data they can retrieve but delaying retrieval until the data is requested.  Data analysis and filtering are done via a separate set of modules called Processors.

We have been using this system for four years and it has been a huge success. The data access implementation has proven to be very easy to use and extremely efficient.  Reconstruction jobs are easy to configure and additional event filters can be added efficiently and robustly -- access to correct data is automatically guaranteed.  The true test of success: physicists have embraced this model for analysis even though the old model is still supported.
\end{abstract}

\maketitle

\thispagestyle{fancy}


\section{INTRODUCTION}

To deal with the vast amounts of data collected, HEP experiments usually write their own data access software.  This software has to be used for a multitude of tasks: calibration, reconstruction, monte carlo generation, and analysis.  In this paper we compare the 'standard' data access system design to the CLEO III design which uses an 'on-demand' mechanism known as lazy evaluation.  In addition, we describe our experience of having used the CLEO III system for the past four years and in particular the response of the physicists who use this system.

\section{STANDARD SYSTEM}
\subsection{Description}
The standard system design~\cite{basf,d0} that most HEP experiments use for their data access software is primarily designed to deal with the task of reconstruction.  For this task, all data objects need to be created for each event that is being processed, since it is the ultimate goal of this task to write all those data objects to persistent storage.

In the standard system design, each processing step is partitioned into its own software Module.  For example, the track finding algorithm and the track fitting algorithm are placed into their own software Modules.  In addition, there are Input Modules responsible for reading objects from persistent storage and Output Modules responsible for writing objects to persistent storage.

To form an actual processing job, the software Modules are run in a user-specified sequence.  When it is a Module's turn to run, that Module executes its algorithm and then places its data into the event.  In addition, if a Module for some reason rejects that event, then the Module can tell the system to stop processing that event (thereby skipping the processing of all Modules appearing later in the sequence) and to restart the sequence with the next event.

\begin{figure*}[t]
\centering
\includegraphics[width=135mm]{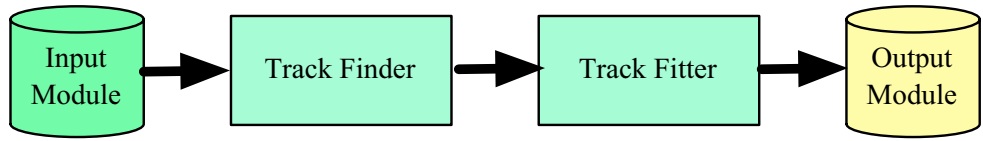}
\caption{Example standard system processing job.} \label{standard_job_example}
\end{figure*}

A simple example of a processing job run using the standard system design is shown in Figure~\ref{standard_job_example}.  In this example four different Modules are used to process an event.  The first Module is the Input Module which reads objects from a persistent object store (in this case it would be objects corresponding to hits that have been calibrated) and then inserts those objects into the event.  The second Module in the sequence pulls the calibrated hits from the event and runs the track finding algorithm which then inserts the found tracks into the event.  The third Module pulls the calibrated hits and the found tracks out of the event, refits the tracks and then inserts the refitted tracks into the event.  The final Module is the Output Module which pulls objects out of the event and writes them to a persistent object store.

\subsection{Critique}
The standard system design has worked well for many experiments for a number of reasons.  One of the main reasons is it has a very simple conceptual model for how a job actually runs.  The job simply runs a series of algorithms that are in a user defined sequence.  This simple conceptual model makes the physicists feel confident that they know how the system works which in turn aids in acceptance of the system.  Another reason experiments use the standard design is that jobs are fairly easy to debug since it is easy to determine which module had a problem.

Unfortunately, the simplicity of the system leads to a number of problems, particularly when the system is used for analysis.  One of the main problems is the physicists using the system must know the inter-module dependencies in order to place them in the correct sequence.  To avoid this problem, physicists often run jobs with many Modules they do not need in order to avoid missing a Module they might need.  Another problem is that optimization of the Module sequence must be done by hand.  For example, reconstruction jobs often have filter Modules which stop the processing of events that contain no interesting physics (e.g. accelerator related backgrounds).  To get the best performance for the reconstruction job, only the Modules creating the data needed by the filter Module should be ahead of the filter Module in the Module sequence.  Since the Module sequence must be explicitly set by the physicist it is up to the physicist to find the optimal Module sequence.  A final problem is reading back from storage is almost always inefficient.  The reason is that all the objects from storage must be created at the beginning of the event processing, since that is where the Input Module is in the sequence, even if the job does not use all objects.

\section{ON-DEMAND SYSTEM}
\subsection{Description}
The on-demand system used by the CLEO~III experiment \cite{cleo3} was primarily designed to be used for analysis batch processing.  The main difference between reconstruction and analysis processing is that in analysis processing not all data objects need to be created every event.  Therefore the on-demand system only creates data objects as they are requested.

Similar to the standard system, the on-demand system breaks the processing task into separate Modules.  But in the on-demand system the types of Modules are further refined into two categories , each with two Module types.
\begin{description}
\item[Provider:] return data when requested.
\begin{description}
\item[Source:] reads data from a persistent store.
\item[Producer:] creates data on demand by running an algorithm.
\end{description}
\item[Requestors:] sequentially run for each new event.
\begin{description}
\item[Processor:] analyzes and filters events.
\item[Sink:] writes data to a persistent store.
\end{description}
\end{description}

Since data objects are only created when requested, data providers register with the system what data they can provide.  This registration is how the system knows which provider to request the data from when another Module requests that type of data.  Since the method used to create a data object (either by reading from a persistent store or by running an algorithm) is activated the first time that data object is requested, the processing sequence is now set implicitly by the order of data requests and not explicitly by the physicist.  Essentially the processing sequence is automatically determined by the system without the need for manual intervention.

While the data providers set what data is available in a job, what is done with that data is determined by the data requestors.  Processors analyze events (e.g. fill histograms or ntuples) and/or filter events of interest by stopping further processing of certain events.  In this system only Processors can stop the processing of an event. The order of the Processors must be explicitly set by the physicist since different Processor orders can be meaningful and only the physicist knows which order is proper for the job.  For example, a physicist could place the event display either before or after an event filter to either see all the events going into the filter or to just see the events that pass the filter.  In contrast, Sinks are always at the end of the processing order and therefore only write out data for events that pass all the filters.

\begin{figure*}[t]
\centering
\includegraphics[width=135mm]{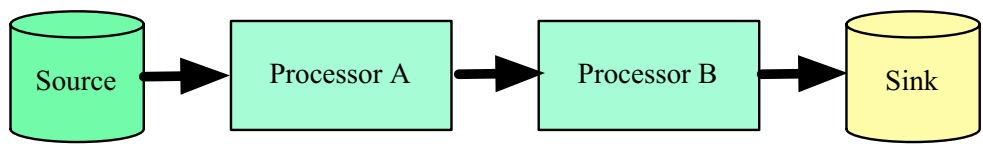}
\caption{Example on-demand system processing job.} \label{on-demand_job_example}
\end{figure*}

Figure~\ref{on-demand_job_example} shows how an event is processed in the on-demand system.  A Source determines the next event to be processed (but does not fill that event with data).  This event is passed to the first Processor in the sequence.  If that Processor does not reject that event, the event is passed to the following Processor in the chain.  This continues until one of the Processors rejects the event, at which time the system goes back to the Source to get a new event, or all Processors have been run.  If no Processor rejects the event, the event is given to the Sinks for storage.  

\subsection{Data Model}

A data access system provides not only a mechanism for accessing data but also a data model that describes how the data is organized.  The CLEO~III data model is unusual because it provides a unified model for all data so, for example, event data and calibration data are treated in the same manner~\cite{c3data-model}.

\begin{figure*}[t]
\centering
\includegraphics[width=135mm]{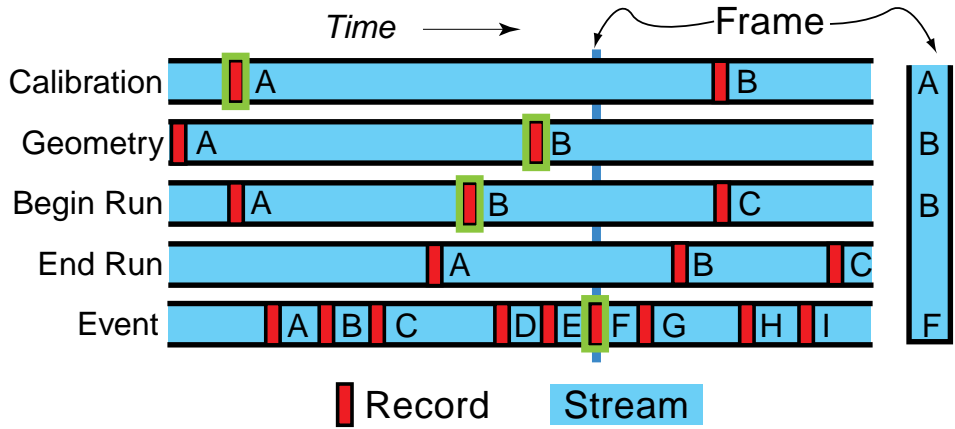}
\caption{The CLEO III data model.} \label{frame}
\end{figure*}

Figure~\ref{frame} gives a graphical view of the CLEO~III Frame/Stream data model.  In this data model, all data items are held in Records.  Data items are grouped into Records based on the 'life-time' of the data.  For example, since both fitted tracks and electro-magnetic showers are only relevant for the instant an event is recorded, both those data items are placed in the event Record.  In contrast, the energy of the beams in the accelerator's storage ring is constant through-out a data run so the beam energy is held in the begin run Record rather than the event Record.  Streams are a time ordered sequence of Records of the same type.  For example, the event Stream holds all the event Records in the proper time order.   The last concept in the data model is the Frame.  A Frame is a collection of Records that describe the state of the detector at an instant in time.  For example, to understand the data in event Record 'F' requires the data at begin run Record 'B', geometry Record 'B' and calibration Record 'A'.

Because all data are accessed in the exact same way, the physicist's learning curve is reduced.  For example, accessing fitted tracks or detector alignment data are done in exactly the same manner.  First the physicist must get the appropriate Record from the Frame.  Second, he must ask the Record to return the data in question.  Also, since the event Record is no different from any other Record, physicists can study any type of data in the exact same way they study event data.  For example, a physicist could analyze the way the alignment of the detector changes with time by creating a Processor that analyzes detector alignment Records rather than event Records.

Since the Frame contains all data relevant for any processing job, the system is designed so that the different component Modules that make up a job can only communicate with each other via the data they have registered in the Frame.  In a sense, this makes the Frame the communication bus for the system.  Using the Frame as the communication bus allows us to avoid explicit dependencies between Modules.  Modules are only dependent on the data objects they use or produce, and not on how those data objects are created.

\subsection{On-Demand Mechanism}

Since the CLEO~III system uses on-demand processing, we need a mechanism to communicate between the data requestors and the data producers.  The mechanism we use employs Proxies.   A data provider registers a Proxy for each data type the provider can create.  These Proxies are placed in the appropriate Record and are indexed by a key.  The key is composed of three different tags:

\begin{description}
\item[Type:] the class type of the object returned by the Proxy
\item[Usage:] an optional compile-time string describing the use of the object 
\item[Production:] an optional run-time settable string.
\end{description}

The two string tags used in the key allow a Record to contain many objects of the same class type.  The Production tag allows us to compare the results of two providers which create data with the same Type and Usage tags.

Physicists access the data via a type-safe templated function call.  E.g.,
\begin{verbatim}
List<FitPion> pions;
extract(iFrame.record(kEvent), pions);
\end{verbatim}

In this example, the first line defines a variable named 'pions' to hold a list of FitPions.  In the second line, the templated 'extract' function is given the event Record (which is obtained from the Frame via the member function 'record') and the 'pions' variable.  The Type tag is determined at compile time based on the type of the 'pions' variable.  The Usage and Production tags are set to their default values.  Once the key is built the extract call asks the Record for the Proxy.  After the Proxy is obtained, the extract call tells the Proxy to run the algorithm to deliver the data.  If the Proxy's algorithm runs successfully, the data is cached in case another request for the data occurs later in the job.  If an error occurs while trying to obtain the data, an appropriate exception is thrown.

\begin{figure*}[t]
\centering
\includegraphics[width=135mm]{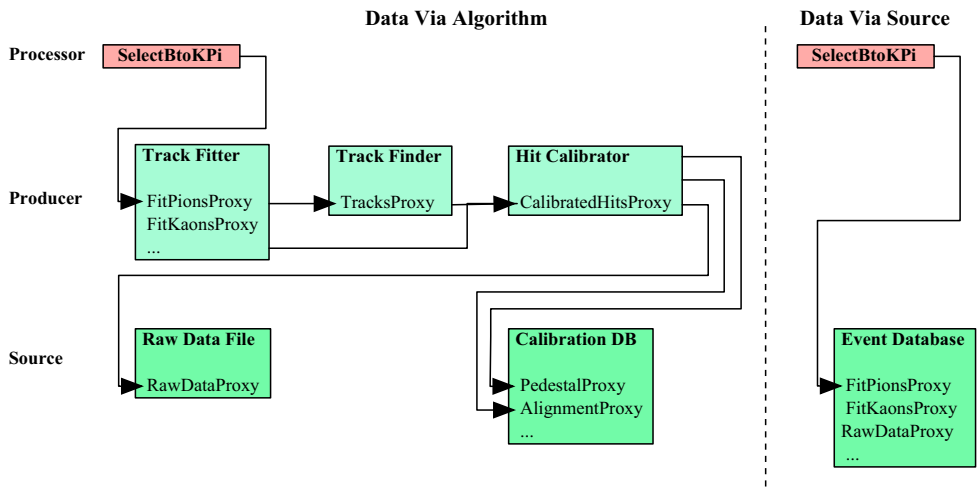}
\caption{Examples of the on-demand system.  On the left PionFits are obtained by running algorithms while on the right they are obtained from the event database.} \label{proxy_examples}
\end{figure*}

Two examples of the on-demand system in action are shown in Figure~\ref{proxy_examples}.  In the left hand example, the Processor SelectBtoKPi gets FitPions by running the full tracking reconstruction algorithm.  In the example, The FitPionsProxy extracts Tracks and CalibratedHits.  The TracksProxy also extracts the CalibratedHits.  The CalibratedHitsProxy then extracts the RawData (which is obtained from a file) and the pedestal and alignment calibrations which are stored in the Calibration database.  In the right hand example, the exact same SelectBtoKPi, run in a different job, now gets its FitPions directly from the event database.  Through the use of dynamic loading, the exact same SelectBtoKPi binary object can be run in the two different jobs without the need to relink the main executable.

\subsection{Critique}

The on-demand design has many positive features.   First, this design supports data access for all HEP jobs: online software trigger, online data quality monitoring, online event display, calibration, reconstruction, MC generation, offline event display and physics analysis.  Second, in contrast to the standard design, the physicist does not need to explicitly set the proper procedural order of the job.  The proper order is determined automatically simply by the order in which data is requested.    Third, the design optimizes access from storage.  At the beginning of processing a Record, a source only needs to say when a new Record is available.  Decoding or retrieval of the data associated with that Record can be deferred until the first request for the data.

However, the added complexity of not explicitly setting the procedural order does make debugging and profiling more challenging.  Good encapsulation of Modules through the Frame mechanism ensures that problems are isolated within a Module and are not from Module to Module interactions.  With such isolation, you only need to know which Module you are in when a problem arises in order to know where to start debugging.  For extremely serious problems that cause a signal to be thrown by the operating system (e.g., a memory access violation), looking at the last few entries in a function call trace-back in a debugger is usually sufficient to isolate the error.  For standard run-time problems we have found that the use of exceptions is critical in understanding the system.  The first implementation of our system used null pointers to data to signify that a problem had occurred during an attempted data access. This lead to a number of problems:
\begin{itemize}
\item it was impossible to know why a problem occurred.  E.g., if FitPions could not be accessed was it because track fitting failed or because no calibrated hits were available?
\item developers had to propagate any error encountered while accessing data needed for their algorithm.
\item the return value of a data access had to be checked to be certain that data was obtained.
\end{itemize}  
Using exceptions solved all of these problems.  Now when an exception occurs, the type of and message in the exception explain the problem.  Additionally, only in blocks of code that catch the exception and continue would it be necessary to check to see if the returned data was valid.  To further aid in understanding what happened when an exception is thrown, we created a stack that holds the present data request chain.  When an exception is caught by the system (because no Module caught the exception), we print a message such as
\begin{verbatim}
ERROR: caught an exception:
"Starting from SelectBtoKPi extracted
[1] type "List<FitPion>" 
     usage "" 
     production ""
[2] type "List<Track>" 
      usage "" 
      production ""
[3] type "List<CalibratedHit>" 
     usage "" 
     production "" <== exception occurred

No data "List<CalibratedHit>" "" ""
   in Record event. 
 Please add a Source or Producer to your
  job which can deliver this data."
\end{verbatim}
and then either terminate the job or skip this event and begin processing the next Record.

\section{WHAT WE HAVE LEARNED}

We have been using a version of this system since September 1998.  During that time we have gained a great deal of experience with an on-demand system.  One of the most important findings is that the on-demand mechanism can be made fast.  We have found that the Proxy lookup takes less than 1 part in $10^{-7}$ of the CPU time on a simple job that processed 2000 events/s on a moderately powerful computer.  Another finding is that cyclical dependencies (a Proxy that winds up extracting its own data) are easy to find and fix.  We only have had one cyclical dependency and it showed up on the first test of the program by causing a stack overflow.  In addition, we have found that we do not need to modify data once it has been created, our Records hold immutable data.  In the cases where we need a new version of the data, we put the new data into the Record with a different key.  One surprising finding is the on-demand system automatically optimizes the performance of reconstruction jobs.  For example, it was trivial to add a filter to reconstruction which removed junk events by using found tracks rather than fitted tracks.  The on-demand system made sure only the algorithms needed for the junk filter are run so that the minimum time is spent processing the event before the filter decides whether to keep it.  We have also found that the best way to speed up analyses is to store many small objects. That way the job only needs to retrieve and decode the data needed for the current job.

Because the on-demand design is so different from the standard design, we have been very concerned with physicists' reaction to the system, particularly since CLEO's previous data access system was a standard design implemented in FORTRAN.  In general, their response has been very positive to the new system.  The programmers of the reconstruction code like the system.  We made it easy for them to get started by having code skeleton generators to make a Proxy, Producer or Processor.  By using the generators, coders only have to write the code specific to their algorithm and not write the code required to work within the system.  In addition, the reconstruction coders found it very easy to test their code since they could easily swap Modules to compare the results of the jobs.  This is extremely easy in our system since we use dynamic loading to load in the different Modules a job uses. Therefore when developers are making changes they only need to recompile their own Module, not relink the entire system.

To ease the transition for our physicists from the old standard system to the new on-demand system, we made sure that they could still program their analysis the 'old way', i.e. all analysis code in the 'event' routine.  However, some of our analysis coders are now pushing the bounds of the system.  These physicists are placing selectors (e.g., cuts for tracks) in Producers.  This allows them to reuse the same binary for many different analyses.  We are even seeing physicists share these selector binaries.  So once selections are done in Producers, the physicists only use Processors to fill histograms or ntuples and to filter events.  Then if a physicist stores the results of her selections into a persistent store, any subsequent pass through the data only requires reading back the selections from storage (and not rerunning the algorithm) and rerunning the Processor.

\section{CONCLUSION}

Our five years of experience with using an on-demand system has shown us it is possible to build such a system so that it is efficient, debuggable, capable of dealing with all the different types of data (not just data in an event), easy to write components, good for reconstruction and acceptable to physicists.

We believe that several reasons contributed to our success.  The first is our use of skeleton code generators, so that physicists only have to write new code, not infrastructure 'glue'.  Second, users do not need to register what data they may request, only what data they provide.  This makes life easier for physicists since data reads occur more frequently than writes.  And third, we use a simple rule for when algorithms are run: if a Producer is added it takes precedence over a Source that can deliver the same data.

\begin{acknowledgments}
This work was supported by the National Science Foundation.
\end{acknowledgments}


\end{document}